# The role of context and culture in teaching physics: The implication of disciplinary differences

Edward F. Redish[*]

[1] Department of Physics, University of Maryland, College Park, MD, USA

## 1. Introduction

The stated theme of the first World Conference on Physics Education is *Context, Culture, and Representations*. This is highly appropriate for an international conference bringing together physics teachers and education researchers from many nations. As an introduction and overview to the conference, I want to talk about how we might begin creating a way of talking about these complex issues that allows us to build our knowledge cumulatively and scientifically.

In any science, there are typically three complementary approaches that support the science – observation (experiment), practice (engineering), and mechanism (theory). Generally, these three approaches intertwine, inform each other, and provide support for each other. I have illustrated them in figure 1 as the legs of a three-legged stool. And as we all know well, the most important leg of a three-legged stool is the one that's missing. In PER we have a strong tendency to focus on observation and practice: how do we see our students behaving and how can we figure out how to teach them more effectively?

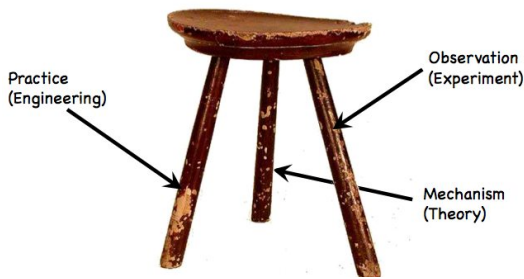

Fig. 1: The three legs supporting scientific knowledge building

The issue of how to build a coherent mental picture (theory) of what happens in a student and a classroom is often the missing leg. While many educational theories exist, they are often narrow prescriptions that provide heuristics rather than frameworks for the development and testing of models that can grow and accumulate knowledge scientifically.[1,2]

Part of the problem is, of course, that human behavior is extremely complex. We cannot expect at this stage to have anything like a complete theory. But it is clear that whatever we do, we will have to consider cognition (a model of how thinking takes place) and socio-cultural environments (a model of how an individual interacts with the context and cultures around her). In section 2 of this paper I give the bare outlines of the beginning of such a framework, the *resources framework*,[3] including three small experiments that you can carry out for yourself to see the validity of the basic principles.

I then show in section 3 an example taken from our studies following real students carrying out real classroom activities in an algebra-based physics class that demonstrates how this theoretical frame-work allows us to describe and model an observed strong *context-dependence* in student behavior by introducing the idea of *epistemological framing*.

---

[*] Address correspondence to: redish@umd.edu

In section 4, I discuss the role of culture. I begin by discussing the impact of diverse scientific cultures on our instruction on non-physics science majors. Currently, the University of Maryland Physics and Biology Education Research Groups are participating in a multi-university multi-disciplinary effort to reform science education for biology majors and pre-medical students. We have held many hours of discussion with faculty in biology and chemistry and have carried out extensive probes into student perceptions about the relations among the disciplines. These have revealed unexpected cultural differences among the sciences, both for faculty and students, that make it challenging for physics faculty to understand how their non-physicist science, technology, engineering, and math (STEM) students interpret our instruction, and make if difficult for our students to connect what they learn from classes in different departments. The section finishes by briefly addressing the important and interesting question of distinct national cultures and how they can play a role in physics education research (PER).

Section 5 talks about the role representations play in physics and how they interact with cultural and disciplinary questions. Section 6 provides a summary and conclusions.

## 2. Talking about thinking:
## A language for discussing context and culture

To understand how to teach students how to learn and understand science we have to understand something of what it means to understand something. It's important therefore for us to find an appropriate level of description for student thinking. We want to follow the basic precept:

> *Everything should be as simple as possible – but not simpler! (Attributed to Einstein)*

What's the appropriate level of description for a system as complex as a science classroom filled with human brains? The human brain is an amazingly complex and flexible device, capable of creating art, science, and culture. In our desire to have something tractable and easy to work with, we have to be careful not to create something *too* simple that does not take into account the full possibilities of the brain's dynamics and creativity.

Despite its great range and flexibility, the brain operates within constraints and structures that have significant implications for our classrooms. To get a sense of this, let's consider three exercises that illustrate some of the basic principles in your own brain.

**Seeing it in your own brain**

The main principles I want to rely on for this talk can be illustrated with three simple experiments that you can do yourself. Try them out before looking at the answers.

> *Experiment 1*

In the first experiment, you are shown 24 words (given in the list shown in the figure at the top of the next page). Look at these words for one minute and try to memorize as many of them as possible.

Don't do anything special or organized[†]: just look at the words and try to remember as many as you can. After one minute, look away and try to write down as many as you can recall.

---

[†] Many of you can construct methods that allow you to remember all these words. This kind of thinking is what we are trying to teach our students to do! To illustrate naïve student thinking, try to do this task without using any highly developed learning skills.

| | | | |
|---|---|---|---|
| Thread | Sewing | Bed | Blanket |
| Thimble | Cloth | Rest | Doze |
| Pin | Sharp | Awake | Slumber |
| Haystack | Injection | Tired | Snore |
| Eye | Point | Dream | Nap |
| Knitting | Syringe | Snooze | Yawn |

Fig. 2: The list of words to try to memorize for experiment 1.[4]

Now look at your list. Check the endnote at the end of this sentence to see if you had either of the two test words on your list.[5] When I give this task to my class, typically more than half of the students put one or both of the test words on their list and are shocked to discover that they weren't there. They were sure they remembered seeing them!

This illustrates a critical principle of memory: that memory is not veridical. It's not accurate like a recording, but rather is "reconstructed" from remembered bits and pieces and plausible "stock items". There is a lot of psychological data supporting this, going back to Bartlett[6] in 1932. More recent support for this result is given in Kotre's popular summary[7] and a modern theoretical interpretation (with support from neuroscience experiments) is presented in Buckner and Carroll.[8]

### Experiment 2

To do our second experiment, you need an internet connection. In this task, a group of six students (shown in figure 3) serve as two teams, one with white shirts and one with black. Each team has a basketball and during the short video they move around quickly, passing their ball among members of their own team. Your task is to see how well you can concentrate by counting the number of passes among the members of the white-shirted team. You have to pay careful attention, since things happen fast!

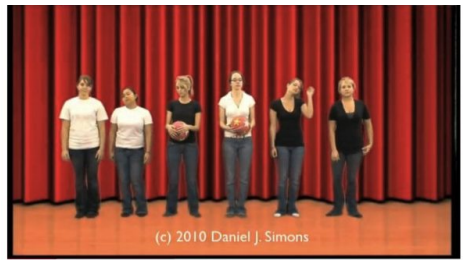

Fig. 3: Daniel Simon's concentration task.

Go to the link http://www.youtube.com/watch?v=IGQmdoK_ZfY and maximize the screen without reading any of the text there (or below) until you are done.[9]

Many people manage to count the number of passes successfully, but fail to see the dramatic events and changes that take place during the clip and are identified at the end. This surprising phenomenon is called *inattentional blindness* – the fact that when you are paying attention to one thing you think is important, you may miss other important things. This is the psychological core of the phenomenon that I refer to as *framing*. It will play a critical part in my interpretation of the role of culture in the classroom.

### Experiment 3

Our third experiment demonstrates that our brains have difficulty in managing tasks of too high a complexity at one time. For this task you will need a partner. Have your partner read you the following strings of numbers and you try say them back in reverse order. So if your partner says "123" you respond "321". Now try it with the following number strings:[‡]

- 4629
- 38271
- 539264
- 9026718
- 43917682

Get the idea? It gets harder and harder and above a certain point it's impossible. Of course you can develop techniques to do this task, but all of my experiments are designed to show that the untrained brain has limitations. This limit on processing capacity has been known for more than 50 years since George Miller proposed his limit of "7 ±2"[10] and is the basis of the important psychological construct of "working memory".[11]

To see that this result has implications beyond this trivial "zero-friction" example, take a look at A. H. Johnstone's 1996 Brasted Lecture.[12] In it, he reports on a chemistry exam on the topic of the mole (Avogadro's number of molecules) set by the Scottish examination board and given to 22,000 sixteen-year-old students. Student success is plotted as a function of the sum of (1) the pieces of information given in the question, plus (2) the additional pieces to be recalled, plus (3) the number of processing steps required. The result is dramatic and shows a sharp drop-off at six pieces of information, consistent with Miller's suggestion. The result is shown in figure 4.

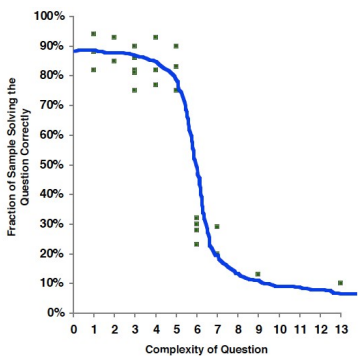

Fig. 4: Dependence of the success of chemistry students on the number of elements in the question.[12]

**Implications**

These little experiments illustrate a few basic principles:

1. Memory is not just recall but is reconstructive and highly dynamic.
2. Selective attention matters.
3. Working memory (what you can hold and manipulate in your mind at any one instant) is limited.[§]

These principles have important implications for the role of context, culture, and the need for the use of external representations. Being aware of our students' spontaneous behavior can help us learn to devise activities to help them learn ways of thinking to overcome them. Of course these don't tell the whole story. For more discussion and lots more references, see refs. 1 and 2.

---

[‡] Thanks to Marjan Zadnik for this example and the reference to Johnstone's work.
[§] The "things" that can be held in working memory and manipulated may be not only single elements but "chunks" – clusters of bits of knowledge that are effectively "compiled" and can be manipulated as a single unit but later unpacked. See refs. 11, 1, and 2 for more discussion.

**The cognitive structure**

The basic principles are just a first step. To figure out how they work in the brain dynamically, we need a more mechanistic picture. I've outlined a model of this based on neuroscience and psychology research in figure 5. Let's imagine that the brain is presented with a straightforward set of data: the perceptual signals associated with holding a cup of Turkish coffee. These include a variety of sensations: (1) *visual* – a pattern of signals arriving on the retinas of your eyes, (2) *haptic* – the sense of touch including the feel, texture, and weight of the cup in your hand, (3) *olfactory* – the smell of the coffee, and (4) *memory* – your knowledge of the cup, including how it tastes, what the effect of the coffee might be on you, and your social knowledge about how and when to drink it – and when to *stop* drinking it so you don't get a mouthful of grounds.

The first step in the way the brain appears to work is that the basic sensory data is processed to create a single coherent perceptual construct and generate some immediate and strong associational knowledge. While it is doing this, it sends signals to the judgment and decision-making part of the brain, the pre-frontal cortex – just behind your forehead. This part of the brain accesses information from long-term memory to decide what to do with the data. This is where your knowledge about the way the world works is brought in. Selective attention (such as in experiment 2) happens here and other perceptions and associations (such as in experiment 1) can now be linked to the original percept. (For more on the details of this process and neurologically explicit examples of it, see Fuster,[13] Bar et al.,[14] and Mesgarani & Chang.[15])

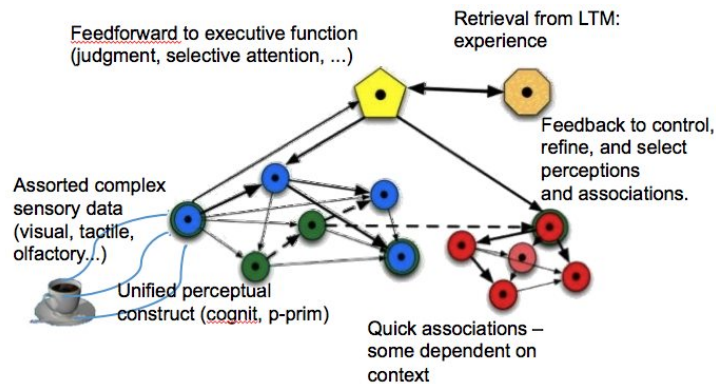

Fig. 5: Dynamical structure of the brain's response to input data.

This model sets us up the structure for the phenomenology I will use to analyze and describe context, culture, and representations: associations and control of those associations.

- *Associations* – Activity in the brain consists of turning on clusters of neurons. These clusters link to other neurons and send out signals that tend to activate (or inhibit the activation of) other neurons, activation of one cluster may induce activation of other clusters leading to interpretation and meaning making.
- *Control* – The feedforward and feedback of signals to and from the prefrontal cortex and long-term memory may activate or suppress the activation of associational clusters.

The control level is where students' assumptions, expectations, and culture draw on their broad knowledge of appropriate behaviors to affect what they do in our classrooms. To understand how to talk about this, let's consider how the behavior of an individual is imbedded in a socio-cultural environment and how this environment affects behavior.

**The cultural structure**

The behavior of any human being is immensely complex. It can be analyzed at scales, ranging from the very small (how many neurons are being activated) to the very large (how does it depend on the presence of highly structured modern technology or the modern nation state). It responds to the individual's knowledge of the human social world, which comes from many sources and scales. I display one way of thinking of this in figure 6. I use a staircase as a metaphor for resolution (or "grain size"): when looking at something while standing at the bottom you can see all the local detail. The higher up you are, the less detail you see – but you are able to discern broader emergent patterns. In a burst of overindulgent nomenclatural enthusiasm I have dubbed this the *cognitive/socio-cultural grain-size staircase*.

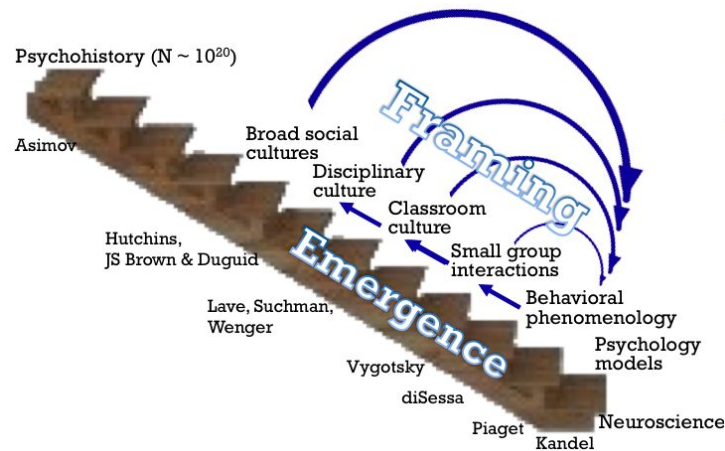

Fig. 6: The cognitive/socio-cultural grain-size staircase.

On the lowest level we see neurons and their functioning – the fundamental matter of which behavior is made up. When we move up a step we ask ourselves about the basic psychological mechanisms of behavior – what they are and how they develop. Another step up takes us to basic behavioral phenomenology – what individuals know about the physical world and how they interact with it. The next step moves beyond the individual and places him or her in the context of a small group. Beyond that, we consider the individuals' relation to the broader local culture of the environment – their knowledge and experience with classrooms and school and their understanding of appropriate behavior in that context. The classroom itself then gets imbedded in multiple cultures – the culture of the discipline being taught and the way schooling is imbedded in the broader culture of the locale – how schooling tends to be viewed by other individuals in the society, how it relates to employment opportunities, how one's position as a member of various subgroups in society affects one's behavior, and so on. Power relations, stereotypes, and other important factors come in at this level. At a step up, we might begin considering the behavior of groups of individuals whose function has to be seen as a group. A software development corporation may have coherent capabilities that no individual possesses[16]; a battleship may know how to navigate but no single individual may have that knowledge.[17]

Each level is emergent from the level below it; the critical behaviors seen at a given level are a result of structures at lower levels, but they may not be easily visible or even discernable in a study of the lower level.

But the important part of this staircase analysis for us as science educators is that the behavior of the individual student in our classroom is affected – often strongly – by the knowledge they bring to the classroom. And their knowledge and perception of <u>all</u> the upper levels of the staircase can play a critical role, serving as control structures for what behaviors students engage in and what they avoid. I show this in the figure as arrows looping back to the basic behavioral level. In order to talk about how this works, I adapt the process known as *framing* from anthropology and sociolinguistics.

**Framing: The interaction of the cognitive and the cultural**

Socio-cultural effects on the classroom have been studied extensively for many decades, but often a critical point is not made explicit. It's not just the socio-cultural environment that matters: *it's a student's perception of the socio-cultural environment that affects that student's behavior*. This requires us to not simply look at the environment and interpret it through our own perceptions, but asks us to consider what socio-cultural knowledge the student brings to our classroom and how that knowledge is used. As in our experiment 2, if our students don't perceive what we have set up for them or asked them to do, it might as well not be there.

The anthropologist Erving Goffmann[18] focused much of his research on the subject of figuring out how people interpret and respond to the social environments they find themselves in from moment to moment. He suggested that people are continually asking themselves the question, "*What's going on here?*" (Though not necessarily consciously.) The answer to that question then controls (again, not necessarily consciously) what behaviors the individual activates. Goffmann referred to the process of answering that question by drawing on experiences stored in long-term memory, as *framing*. The concept has been further developed in sociolinguistics[19] and in other fields as well.[20] For an extended discussion of how it applies in physics education, see Hammer et al.[21]

Framing can have many components: for example, affective (How will I feel about this class?), social (Who am I going to interact with and in what ways?), and epistemological. This last is particularly important for a science class where we are trying to build a students' knowledge. I consider *epistemological framing* to be the process that generates each individual's answer to the questions:

> *What is the nature of the knowledge we are learning in this class and what do I have to do to learn it?*

The concept of framing matches well with the more recent developments I cited above in neuroscience showing that sensory data is put in context by passing signals through the pre-frontal cortex, which draws on long-term memory to adjust attention and to activate knowledge and decisions of how to behave. Framing is what you did when you focused your attention on the passes in experiment 2 and as a result wound up not seeing other elements that were interesting and possibly important. In that case, in my instructions I encouraged you to frame the task as a concentration one, which encouraged you to ignore (or even suppress) everything else that might be happening. But problems occur in a classroom when students bring in their own expectations that may result in their ignoring messages that you think you are sending explicitly; expectations like, "I know how a science class works. I don't have to read all these pre-class handouts."

## 3. Context

We now have a language to talk about how students respond to context in our classrooms. Let's consider a specific detailed example from the work of Brian Frank.[22] This is taken from a lesson in a class in Introductory Physics (algebra based) at the University of Maryland. The classes are taught in fairly traditional structure, with three hours of large lecture (N ~ 200) per week, one hour of small-group recitation (N = 24), and two hours of laboratory. The population was largely life and health science majors. The class had been modified to place more emphasis than usual on epistemology – How do we know? Why do we believe? (The modifications are described in detail my paper with David Hammer.[23]) The specific example I want to describe occurs in the recitation section, which were run as Tutorials.

**Tutorials**

The term "Tutorial" has a wide variety of meanings in different countries around the world. But when a physics education researcher in the USA talks about "Tutorials" they almost always mean "University of Washington-style Tutorials." These are lessons developed over many years by Lillian McDermott and her colleagues at the University of Washington.[24] These lessons are carefully tested and refined through multiple cycles of research, curriculum development, and instruction. Tutorials focus on well-defined common student difficulties and typically help students understand how to develop qualitative reasoning.

Students doing these lessons work in groups of 3-5 facilitated by a teaching assistant (TA) – ideally about 1 TA per 15 students. The assistant is trained to understand what difficulties the student can be expected to encounter and is taught to encourage the student to explore and discuss their own ideas. (For a detailed description of UW Tutorials and references to their research papers on Tutorial development, see chapter 8 of my book, *Teaching Physics with the Physics Suite*.[25])

Our example is drawn from our modification of an early lesson in kinematics.

**An example**

For the lesson on velocity, we use a standard device (shown in figure 7). A long thin paper tape ("ticker-tape") is attached to a low friction cart (shown at the left) and run through a "tapping device" that taps a sharp point onto the tape through a piece of carbon paper at a fixed rate. The cart is allowed to accelerate slowly down a long ramp and the tapping device creates dots on the tape whose spacing indicates the cart's speed.

The tape is then cut into segments of six dots each. Since the cart accelerated slowly, 6 dots (representing about two-tenths of a second) look as if they are representing a constant speed. Each student receives a segment as shown in figure 8 below.

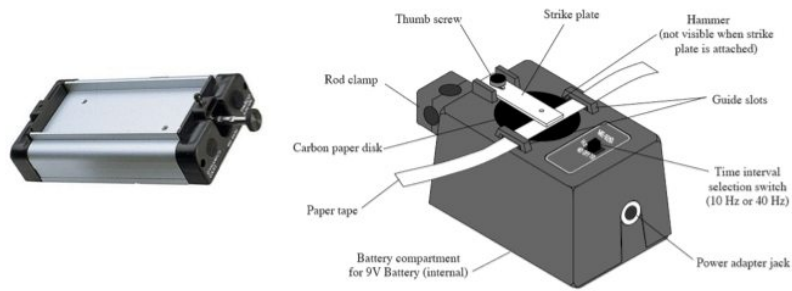

Fig. 7: Pasco low-friction cart and ticker-tape tapper.

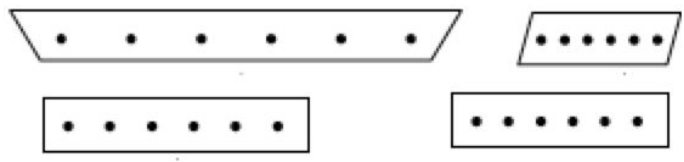

Fig. 8: Samples of ticker tape given to the students in Tutorial.

The first question the students are asked in the lesson is: "How does the time taken to generate one of the short segments compare to the time to generate one of the long ones?"

Since the marking device taps at a fixed rate, the answer is trivial: since they each have six dots, they each took the same amount of time to make. Interestingly, this is not what we saw the students say in our videotapes of the lessons. Here are some transcripts:

Group 1
    S1: Obviously, it takes less time to generate the more closely spaced dots.

Group 2
    S2: (Reading) "How does the time taken…" It's shorter! (Huh!)
    S3:  Yeah. Isn't it pretty much –  The shorter ones are shorter.

Group 3
    S4: (Reading) "The time taken to generate one of the short segments…"
    It's shorter!

Group 4
    S5: Well it takes less time to generate a short piece of paper than it does a long one. (pause) I would assume. (pause) I don't really know how that thing works. [The last two comments are ignored by the rest of the group.]

It's quite dramatic watching one group after another give the same obviously incorrect answer, without hesitation and mostly confidently. This looks suspiciously like a standard "misconception". But the last example we quoted gives a hint as to what's going on.

If we go a bit further into the video, we find the next question in the lesson shifts the context for the group. The result is that they bring a different approach to bear. They are asked, "Arrange the paper segments in order by speed. How do you know how to arrange them?" Here's a typical response from one of the groups.

    S1: Acceleration! It starts off going slow here,[pointing to a short segment] then faster, faster, faster [pointing to a long segment].

    S2: No, no! Faster, then slower, slower, slower! This is slow.[pointing to a long segment]

    S1: When it gets faster it gets farther apart. That means the paper's moving faster through it. [gestures] So it's spaced out farther.

    S2: Wait. Hold on. [gestures to TA]

    S2: [to TA] Is the tapper changing speeds or is the paper moving through it changing speeds?

    TA: The tapper always taps at the same speed.

    S1 and S2 [together and pointing at each other]: Ahhh!

We saw this again and again. At the beginning the students gave a quick answer – longer tapes take longer times, shorter ones take shorter. Just a few minutes later, the light dawns and they all get it right.

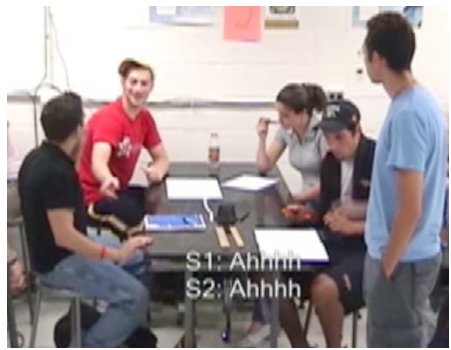

Fig. 9: The light dawns.

What is changing when the students in these groups shift their behavior in response to a (very slightly) changed context? I suggest that the easiest way to describe what is happening is as *epistemological reframing*.

**Is it a misconception?**

In PER we often have a tendency to refer to common errors that students bring into the classroom as "misconceptions". I don't have a problem with this, but I would like us to take a fine-grained view of them. I define a *misconception* as *a student error that is commonly and reliably activated in a given context*. This gives us more flexibility and encourages us to understand what is happening in detail – to consider how the student is responding to the context – rather than using the term to close off further consideration of what the student is actually bringing to the task. Misconceptions can have structure, not just be an irreducible gallstone that needs to be removed. Misconceptions can be robust and hard to undo, but sometimes they are created on the spot and are highly context dependent.

In our example, at the start of the lesson many students have the epistemological assumption that they will be able to generate a correct answer by simply looking at the question and drawing the most immediate and natural response – essentially by an immediate association without carefully considering the mechanism of what is happening ("one-step thinking"). What they get is a *phenomenological primitive* (p-prim): "more is more", which they map in this situation into "longer tape takes longer time".[26,1] This feels right to them and they move on.

The misconception in this case isn't actually a misunderstanding about the nature of velocity; rather it's a *framing error* that is common and reproducible in this context. The misconception here is epistemological rather than conceptual: students assume that the answer can be generated directly by fast thinking without any careful consideration of the mechanism. The student in group 4 expressed a framing caveat: essentially, "We might have to consider the mechanism here." The later questions on the worksheet can't be answered without considering the mechanism, so a frame shift was needed. This led the students to go back and reconsider (and correct) their answer to the first question.

For most physics teachers it will be a surprise that our students might "miss the gorilla in the classroom" and assume they didn't need to think about the mechanism of what's happening – especially since the lesson begins with the TA describing the mechanism! But selective attention can cause them to not only focus on particular aspects of a task but to ignore aspects their instructors might consider natural and critical.

Epistemological framing – what the students think is the kind of knowledge they are seeking and what they think they have to (or are allowed to) use to get it often plays an important role. If we ignore the issue of epistemological framing, we might misinterpret where a common student problem lies and have trouble creating an effective lesson – or fail to understand *why* a particular lesson is effective. (For more examples, see my Varenna lecture.[1])

Here's the takeaway message:

> *Student responses don't simply represent activations of their stored knowledge. They are dynamically created in response to their perception of the task and what's appropriate. As a result, their behavior may have a complex structure. The (often unconscious) choices they make as to how to activate, use, and process knowledge are often determined by social and cultural expectations (framing).*

## 4. Culture

**The disciplines**

The students in the example in the previous section brought a rather local expectation into their classwork – that they could get the answers without thinking about the mechanism of what's happening.

This is likely to come from one step up in the "cognitive/socio-cultural grain-size staircase" – from their experience in the culture of other classes, particularly their previous science classes. In some cases, difficulties arise from additional steps up. In my next example, disciplinary cultures play a dramatic (and surprising) role in reforming an introductory class for biology and health-science majors.

### *Teaching physics is primarily in service to other disciplines*

Physics departments often focus their concern about teaching on their majors. These students are of course important; they are our intellectual progeny and the future of our discipline. But we may tend to forget that physics teaching at the university level is primarily in service to other larger disciplines. In the Physics Department at the University of Maryland we have a total of about 250 undergraduate majors and a comparable number of PhD graduate students. But every semester we teach nearly 1000 engineers and 1000 biology majors in our introductory service courses. A similar pattern can be found at most large research universities in the USA, and a brief email survey of my friends and colleagues in other countries indicate that this is widespread around the world. This is a result of the structure of science today. Figure 10 shows the distribution of PhD scientists in the USA in 2006. Physics is the small blue bar – smaller even than geosciences. Our primary service lcients are biology (the largest bar in green on the left) and engineering (the red bars).

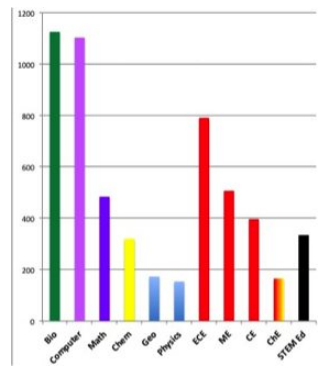

Fig. 10: Thousands of PhD scientists in the US by field (2006).[27]

Many universities in a number of different countries have seen significant cutbacks (or even eliminations) of their physics departments as a result of losing students. How can this be? Don't the engineers and biologists need to study physics? It's basic!

Unfortunately this is an indication that our perceptions of our profession and teaching may be culturally biased. At a recent conference on interdisciplinary science and math education, I had given a talk on the topic of creating a "Physics for Biologists" class.[28] Here are two quotes from a physicist and a biologist later heard and transcribed by my grad student, Ben Geller:

> *Physicist:* This whole 'physics for biology' idea makes me very uncomfortable. What's next? 'Physics for mechanical engineers' or 'physics for electrical engineers'? Where does it end? I could see maybe having a physics class for all students and then having a few tailored recitation sections where students focus on applications to their various fields, but I'm uncomfortable with 'physics for X' as an idea. We should be conveying how we view physics to everyone.

> *Biologist:* I guess the physics for biologists idea may be a step in the right direction, but for it to be useful it has to go much further and be entirely revamped. It has to be very narrowly focused on those ideas that biologists see as essential, not just removing a few topics…. Unfortunately, physicists generally have a profound ignorance about biology, so I'm not sure they are the right folks to be doing it. I can teach the relevant physics myself.

This suggests a dramatic – and chilling – culture gap between the perceptions of a physicist and a biologist about how to teach physics for service students. These comments are not unusual or unique. I have heard comparable ones from many physicists and biologists. If we are to keep teaching physics to biologists (and engineers) we need to understand their culture – and their perceptions of physics and how it is of value to them.

As a result of a recent opportunity to participate in a national reform project, I have begun to learn something about disciplinary cultural differences, their impact on our instruction, and how these differences are viewed by the biology faculty (gatekeepers to our classes) and by our biology students.

### *Calls for change from the biologists: The challenge*

For nearly a decade the biology community has been calling for an upgrade of undergraduate instruction for biologists.[29] In 2009, Association of American Medical Colleges (AAMC) working with the Howard Hughes Medical Institute (HHMI) published *Scientific Foundations for Future Physicians*[30] – a call for rethinking education for biologists and pre-medical students in the US to bring in more and better coordinated science – biology, math, chemistry, and physics, and to focus on scientific skills and competencies.[31]

The result was that HHMI funded *Project NEXUS*: the National Experiment in Undergraduate Science Education, a 4-year, 4-university $1.8 M project.[32] At the University of Maryland, College Park (UMCP) we have opened an interdisciplinary conversation to create a physics course designed to meet the needs of biologists and pre-health-care-professionals.[33] We've put together a team of nearly 40 professionals, including physicists, biologists, chemists, and education researchers both on and off campus. Over the past two years we have held hundreds of hours of interdisciplinary conversations and negotiations among subgroups of this team. We quickly discovered that creating an interdisciplinary physics course that meshed with what was being taught in biology and chemistry and met the needs of life science majors was not going to be simple.

It turned out there were significant cultural differences between biologists and physicists. Biologists saw most of the traditional introductory physics class as useless and irrelevant to biology – and the claim made by the physicists, "We can apply physics to biology examples," as trivial and uninteresting. Physicists saw a coherent structure with no room for change.

After much discussion and negotiation, we came to a better understanding of what it was the biologists needed and how the disciplines perceived the world and their science differently.

### *Culture differences between physics and biology*

What we have learned from our extensive interdisciplinary conversations is that for us to meet the needs of biologists in learning physics, there is much more than changing the table of contents and the prerequisites. Each scientific discipline brings broad cultural assumptions, approaches, and epistemologies that are unique and strongly affect the way that both faculty and students frame the activities in a science class. Here is a list of some of the characteristics that we found were distinctive in physics and biology with an emphasis on how the introductory classes are treated.

*Physics: Common cultural components*

- Introductory physics classes often stress *reasoning from a few fundamental (mathematically formulated) principles*.
- Physicists often stress building a complete understanding of the *simplest possible (often abstract) examples* – and often don't go beyond them at the introductory level.
- Physicists *quantify* their view of the physical world, *model with math*, and *think with equations*.

- Physicists concerns themselves with *constraints* that hold no matter what the internal details. (conservation laws, center of mass, ...)

These elements will be familiar to anyone has ever taught introductory physics. What is striking is that we usually do not articulate what we are doing – and <u>none</u> of these elements are typically present in an introductory biology class. Biologists have other concerns.

*Biology: Common cultural components*

- Biology is *irreducibly complex*. (Oversimplify and you die.)
- Most introductory biology is *qualitative*.
- Biology contains a critical *historical* component.
- Much of introductory biology is *descriptive* (and introduces a large vocabulary)
- However, biology – even at the introductory level – looks for *mechanism* and often considers micro-macro connections.
- Biologists (both professionals and students) focus on *real examples* and *structure-function relationships*.

These issues don't match well with what we tend to do in intro physics. Though we do focus on mechanism, it rarely is explored at the atomic or molecular level. The demand for realism and structure-function relationships was a particular sticking point for us. The physicists on our team often found an example or explanation "cute" or "enlightening" if it helped explain a relationship. The biologists mostly were uninterested in such examples unless they could see how it had implications for real-world examples. Many of our biologists considered traditional physics examples, such as the simple harmonic oscillator (mass-on-a-spring), irrelevant, uninteresting, and useless until we were able to show its value as a starting-point model for many real-world and relevant biological examples. This required making it clear from the first that a Hooke's law oscillator was an *oversimplified model* and illustrating how it would be modified for realistic cases.

### *Restructuring Introductory Physics for the Life Sciences*

As a result of our discussions and negotiations we proposed to change both the culture and the content of the class so as to make the value more obvious both to biology faculty and to biology students. Here are some of the "cultural guidelines" we have chosen.

- Organize the course and select examples so that both biology faculty and students feel that it has obvious value for upper division biology courses.
- Do not assume this is a first college science course. Make biology, chemistry, and calculus pre-requisites.
- Do not assume students will have later physics courses that will "make things more realistic". Explicitly discuss modeling and the value of understanding "simplest possible" examples.
- Choose different content from the traditional by including molecular and chemical examples and topics of more importance to biology.
- Maintain the crucial components of "thinking like a physicist" – quantification, mathematical modeling, mechanism, multiple representations and coherence (among others).

Physicists often assume that the content of an introductory physics course is almost all "privileged" – you have to do it all to get a start in physics. What is often missed is that the standard content is not a complete introduction; it is already a selection. Our current selection tends to favor items that can be done mathematically completely and simply. Topics that are of great importance in biology – such as motion of and in fluids, diffusion, and electrical properties of matter – are suppressed; I suspect in part because a full mathematical treatment of these topics lies at the graduate level. But these topics are needed – and used! – in introductory biology and chemistry classes. We have decided that we can do something useful with these topics by including some phenomenology while still maintaining the cru-cial components of "thinking like a physicist." As a result, we are attempting to include significant treatments of the following topics in our class.

- Atomic and molecular models of matter
- Energy, including chemical energy
- Fluids, including fluids in motion and solutions
- Diffusion and gradient driven flows
- Dissipative forces (drag & viscosity)
- Kinetic theory, implications of random motion, and a statistical picture of thermodynamics

These topics are difficult and cannot be done without considerable effort (and some needed education research). As a result, some traditional elements have to be suppressed. After much discussion we have decided to eliminate or dramatically reduce our coverage of the following elements.

- Projectile motion
- Universal gravitation
- Inclined planes, mechanical advantage
- Linear momentum
- Rotational motion
- Torque, statics, and angular momentum
- Magnetism
- Relativity

Some of these decisions are quite painful to someone (like myself) who has been working on finding good ways to teach some of these topics for many decades. We have had to think carefully about *why* we felt strongly about each topic. For example, we concluded that projectile motion and inclined planes had significant value as a vehicle for teaching students about vectors and components, but little value as core models for anything a biologist would use. As a result, we chose to reduce these topics significantly and replace them by an early treatment of static electrical forces (often in molecular examples) with careful training with vectors.

Note that for almost every topic that we cut back on, we could imagine practical applications for biologists or medical professionals that relied on one or more of them. But one cannot cover everything without becoming so superficial that our students only learn words, not ways of thinking. We decided that most biologists needed to be able to think about and understand basic biological mechanisms more than they needed to understand how their tools and measuring devices (that are treated as "black boxes" by most professionals) work.

Bridging the culture differences between physics, biology, and chemistry has been challenging. But even as we make progress in connecting at the faculty level, how our students respond to our attempts to bridge the cultures of the scientific disciplines also needs to be understood.

### *Student attitudes toward interdisciplinarity*

From each level of their experience with a discipline – small group interactions, STEM classes, broader school experiences – students bring control structures (framing) that tell them what to pay attention to in the context of activities in a science class. Their framing of an activity affects how they interpret the task and what they do.

We have studied student responses to interdisciplinary activities combining physics and biology in two contexts: in an organismal biology course that is attempting to use a principle-based approach that includes a lot of physics, and in our NEXUS physics course that is attempting to teach physics principles in a way that uses authentic biological contexts. In both cases we have measured student responses and attitudes in a variety of ways. There is not space to go over these observations in great detail, but I will point out two examples that illustrate how our cognitive / socio-cultural / framing analysis helps us understand some of the often surprising issues that arise.

*Problems with using physics in a biology course*

Ashlyn[**] is a student in the Organismal Biology course mentioned above. In one lesson, the instructor produced the implication of Fick's law: that the distance something diffuses (in one dimension) is proportional to the square root of the time. Later in the class, Ashlyn made the following comment in an interview.

> I don't like to think of biology in terms of numbers and variables.... biology is supposed to be tangible, perceivable, and to put it in terms of letters and variables is just very unappealing to me.... Come time for the exam, obviously I'm going to look at those equations and figure them out and memorize them, but I just really don't like them.

> I think of it as it would happen in real life. Like if you had a thick membrane and tried to put something through it, the thicker it is, obviously the slower it's going to go through. But if you want me to think of it as "this is $x$ and that's $d$ and this is $t$", I can't do it.

Actually, she "can" do it, because she also took my physics class (previous to our reform to match it to biology) and did very well. I believe this is clearly a framing problem. Based on her experience and expectations, she sees reasoning with mathematics as unnecessary – and even distasteful – in biology.

Interestingly enough, this is not the end of the story.[34] Later in the interview, Ashland got excited about an exercise later in the class in which the implications of surface to volume ratio for biology were explained mathematically. A small wooden horse was constructed of a few blocks of wood and supported with dowels for legs. It stood quite nicely. A second horse was then produced in which every dimension was scaled up by a factor of 2. When placed on the ground the legs broke, unable to hold up the extra weight.[††] This larger wooden horse is shown in figure 11.

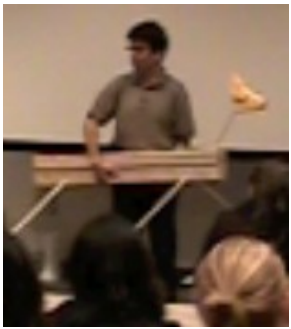

Fig. 11: Jeff Jensen demonstrating his scaled up wooden horse in Org Bio.

Here is how Ashland described her response to this activity:

> The little one and the big one, I never actually fully understood why that was. I mean, I remember watching a Bill Nye episode [TV science program in the US] about that, like they built a big model of an ant and it couldn't even stand. But, I mean, visually I knew that it doesn't work when you make little things big, but I never had anyone explain to me that there's a mathematical relationship between that, and that was really helpful to just my general understanding of the world. It was, like, mindboggling.

---

[**] All names are gender-indicative pseudonyms.
[††] The weight goes up like the cube of the scaling factor (X8), but the strength of the dowel-legs only goes up by the cross-sectional area of the dowel, which scales like the square of the scaling factor (X4).

This pair of statements leads us to an interesting take-away message.

> *Biology students bring cultural/disciplinary expectations to their classes that lead to framings of activities that may get in the way of trying to create interdisciplinary instruction – but it can be context dependent. If the activity is perceived as doing work for them, students can reframe their view of what is going on.*

*Problems with using biology in a physics course*

During the first trial of the reformed NEXUS class in the 2011-12 academic year, we made an attempt to deal with some of the issues that had been raised in our negotiations with the biologists and chemists. One problem that they identified was that students were highly confused about chemical binding. The biologists often used the language of "energy stored in chemical bonds." The chemists (and physicists) were uncomfortable with this language since a "bond" implies a negative energy and that energy has to be put in in order to sever the bond. A critical example is the *hydrolysis of ATP*. The molecule adenosine triphosphate (ATP) has a weakly bound phosphate cluster. In a watery environment a small amount of input energy can break this bond; the phosphate can then bind with water, which forms a strong bond. As a result, significant amount of energy can be made available to do a variety of kinds of biologically relevant work. This reaction is fundamental to biology and ATP is often referred to as *the energy currency of the cell*. The process is illustrated in figure 12.

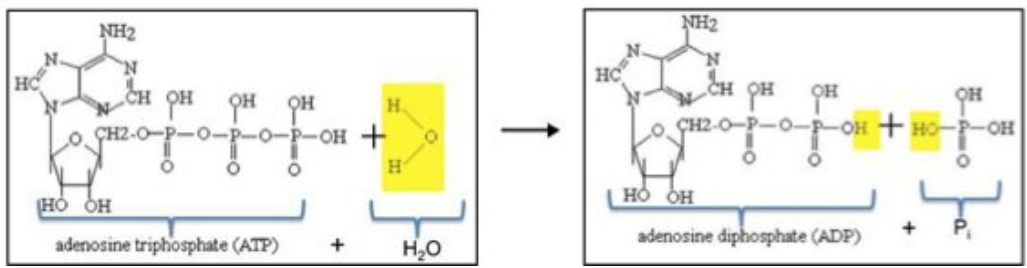

Fig. 12: The chemical reaction hydrolyzing ATP that provides biological systems energy for a variety of essential processes.

Chemistry education researchers often identify as a "misconception" that students assume "energy is stored in the ATP bond" whereas really the energy comes from going from the weaker ATP bond to the stronger OH-P bond.[35]

After going through a series of activities that bridged from everyday situations (a skateboarder in a dip) to simple chemical examples, we tried a question from W. Galley's paper in the chemistry education literature (figure 13) in a quiz early in the second term of the class.[35]

---

An O-P bond in ATP is referred to as a "high energy phosphate bond" because:
(choose all correct answers.)

        A. The bond is a particularly stable bond.

        B. The bond is a relatively weak bond.

        C. Breaking the bond releases a significant quantity of energy.

        D. A relatively small quantity of energy is required to break the bond.

---

Fig. 13: A question probing the bonding misconception from the Chemistry education literature. (Ref. 35)

Answers B and D are considered correct and answer C is considered representative of a serious misconception.

Our students did a bit better than the chemistry class reported by Galley – but not much. In our class 79% chose the "misconception" C compared to 87% in the chemistry class; and 47% chose the "correct" answer B compared to 31% in the chemistry class. However, there was an interesting result. Almost 1/3 of the students gave B and C as answers – apparently contradicting themselves. Why?

A possible answer appeared in an interview with Gregor, a student who joined the class in the second term and was often explicitly metacognitive about his thinking. Gregor chose BCD for his answers and explained his choice this way.

> I put that when the bond's broken that energy is released. Even though I know, if I really think about it, that...you always need to put energy in ... to break a bond. Yeah, but -- I guess that's the difference between how a biologist is trained to think, in like a larger context and how physicists just focus on sort of one little thing.... I answered that it releases energy, but it releases energy because when an interaction with other molecules, like water primarily, and then it creates like an inorganic phosphate molecule that...is much more stable than the original ATP molecule.... I was thinking that [in the] larger context of this reaction [it] releases energy.

This is a very interesting explanation. Gregor explicitly describes his loss of points on the quiz as a framing error. When looked at it from a physics framing, one assumes that one should isolate the molecule and talk about it as if it were in a vacuum. When considered however from a biological framing, the molecule is always in water so the availability of water molecules – and the second part of the reaction – can be taken for granted. In this framing of the context all three of Gregor's answers can be seen as correct.

This last example pulls together a number of the threads we have been discussing through this paper. The disciplines often make different epistemological assumptions so that a given context may be seen in different ways depending on what framing is used. This leaves us with a take-away message for this section.

> *In considering students' responses in interdisciplinary situations (indeed, in ANY situations), we have to be aware of possible framing differences that arise out of the differences between disciplinary cultures.*

**International Studies**

Disciplinary cultures are not the only steps in the cognitive/socio-cultural grain-size staircase that affect how our students perceive activities in our classroom – how they frame the activities and behave accordingly. The broad cultural elements we all learn from being part of a community – a profession, a nation, a family, a religion – plays a role in how we interpret what we see and do. At an international conference it is particularly appropriate for us to consider what we can learn from international and intercultural comparisons.

Physics education research and science education has a long history of international comparisons. The TIMSS, PISA, and other studies have received international attention. The abstract volume for this conference contains many multi-national comparative studies. While these are of considerable interest, what is of even more interest is the following:

> *When comparative international studies find significant differences between comparable populations in different countries, how do we figure out what is responsible for those differences?*

Until we have answered that question, we can't really tell what a nation or school district might need to do in order to improve their teaching and learning. While there have been some attempts to explain international and cultural differences, my sense is that most studies are still at the stage of documenting the differences and don't go much beyond speculation in considering what causes the observed differences.

Let me briefly cite one example of a study that our research group has just begun. In this study we try to use the variability of situations from one country to another in order to probe variations that would be difficult to examine in a single school system.

One issue that arises when considering students' epistemological framing is, "Where and when do students develop frames and how easy is it to 'resurrect' a long-unused frame?"

One of the big problems in implementing Tutorial classes such as described in section 3, is that students often bring serious epistemological misconceptions to the class. They have had *so* much experience with classes where all that mattered was the answer, that they have a hard time focusing on reasoning – why we choose a particular answer. Many have had high school physics classes in which plugging numbers into poorly understood equations sufficed to earn a solid A. As a result, they have difficulty framing the Tutorial as an activity that will contribute significantly to their physics achievement. (This despite extensive data that show much stronger learning with Tutorials than with traditional problem solving recitations.[36]) The result is often discontent and serious resistance. Making Tutorials work in the US often requires careful effort, training TAs to understand the challenge and helping them to learn how to help students make the transition to a more effective epistemological framing of the worksheet activities.[37]

Many of our students at the University of Maryland appear not to have had significant experience in qualitative groupwork in which their own ideas were valued. The result is often significant initial resistance to the kind of activities in Tutorials. Would the situation be different if they already possessed similar experiences and had available an epistemological framing of such activities that they were comfortable with?

It would be difficult for us to find a significant sub-population among our students in the US who have had such experiences. However, different national instructional models allow an exploration of this issue. In Japan, students often are exposed to groupwork in which their own ideas are explored and valued during elementary school. But in middle and high school, high-stakes testing drives the educational model towards more rote learning, drill and practice, and straight lecturing.[38] A graduate student in my group, Mike Hull, wondered whether students in Japan would respond differently to Tutorials than American students and whether or not they would perceive their elementary school experience as helping them adjust to the new environment.[39]

In the Spring of 2011, Mike visited Gakugei University in Tokyo, where he helped Professor Uematsu translate and implement Tutorials in a class of 140 undergraduate pre-service teachers. He did extended interviews with 28 of these students. He found that students took to the Tutorials immediately, without the resistance observed in the US. There is evidence to suggest that many of the students were able to reach back to their elementary school experience and activate an epistemological frame that expected them to interact with each other and use and evaluate their own ideas. One student, when asked why it was so easy to adjust to Tutorials, responded,

> Even though so many years have passed in middle and high school where we were being taught uni-directionally by a teacher, even though we took those classes, the chatting, talking, and solving problems together that we did in elementary school was fun. Talking with people about things that you know, and if that person knows something you don't, he can teach you… since we know the importance of that, we quickly got used to [Tutorials] I think: we have experience from elementary school.

Here's the take-away message:

> *Inter-cultural comparisons provide us with extraordinary opportunities for carrying out "experiments" that could not be done in classrooms in a single culture. These experiments may help us to better understand the developmental trajectory of the epistemological frames students bring to our classes.*

## 5. Representations

The last of the three topics in this conference's theme is *representations*. This fits in extremely well with the issues of context, culture, and with our model of thinking. If you tried experiment 3, you might have been surprised at how quickly your brain ran out of processing and storage space. After all, as a physicist you are likely to have generated arguments and mathematical derivations that ran over many pages – and it felt like you could hold it all in your head at once. But this is where external representations come in. Off-loading cognitive information onto external visualizations allows us to create much more complex reasoning than unassisted working memory can handle. The brain can do fast switching, so having things represented externally allows us to "roll in" and "roll out" knowledge and make connections that are otherwise too much for us to handle. The external elements essentially become a part of our cognitive processes.

Since physics makes so make effective use of so many different kinds of representations, there are numerous studies of how students interact with them and how to help students frame them as coherent (like with the cup of Turkish coffee, merging multiple perceptions to create a single sense of the phenomenon). I don't have much space to discuss this, but I want to make two important points. First, that our use of external representations is woven deeply into the culture of physics, and second, that representations are strongly cultural. Disciplinary traditions and assumptions about representations can lead to conflict in trying to create an interdisciplinary approach to teaching the various sciences to a single population of students.

Recognition is much easier than recall. As a result, we can think effectively using external representations. My favorite example of this is using computers. In my house, both my wife and I create PowerPoint presentations for our work. Sometimes, I know how to do something that she doesn't and I'll be called on to explain. Often, I can do the task but I can't tell her how to do it. I have to sit down at her computer and show her. The problem is that I recognize which menu I will need to use and, when it opens, I will recognize which item I need to choose. Then I recognize what I have to enter in the dialog box that appears. But without interacting with the program itself, I don't remember what the steps are. I don't know how to carry out the task, but the program and I together know how to do it.

In physics, we often "think with equations," using external symbolic representations not just to calculate, but to organize and provide easy access to a large amount of qualitative and conceptual knowledge. Here's an example (an excerpt from the text materials for our NEXUS class) discussing the knowledge represented in the equation expressing Newton's second law, externally represented in mathematical form as:

$$\vec{a}_A = \frac{\vec{F}_A^{net}}{m_A}$$

Fig. 14: Newton's second law represented mathematically.

> Each bit of this seemingly simple-looking equation codes for activating bits of conceptual knowledge about motion.
>
> 1. *a* -- The thing on the left of the equation is the acceleration. To understand that, we have to understand the whole array of specifying an object's position (coordinates) and how that position changes (derivatives, velocity, acceleration). This means (for motion in one dimension) we need the definitions of velocity and acceleration as the derivatives of position and velocity respectively.
>
> It's important to note that the acceleration is written *on the left*. We do this to remind ourselves that *it's the forces that cause the acceleration rather than the other way* *around.* Though of course if we know the acceleration and mass we can find the net force. [Students tend to think of an equation as the way to calculate the thig on the left.]

2. *A* -- Each of the variables has a subscript labeled by which object we are talking about. This reminds us that a fundamental assumption of the Newtonian framework is that we best understand what is happening by considering individual objects and figuring out what influences are acting on them. Each object we consider will have its own Newton 2 tion. The subscript *A* on $F^{net}$ reminds us that it is the forces that the object feels that controls its motion. (The forces it exerts have effects on the motion of the objects it exerts them on.)

3. *F* -- To interpret this we need to understand that it is the interactions with other objects that cause the object we are considering to change its motion (accelerate). And we need to understand how this force is quantified by an operational definition.

4. *net* -- This little superscript holds a lot of conceptual ideas. First, that it is the (vector) *sum* of the forces that an object feels that results in its acceleration. Each individual force does not produce an individual acceleration. When we break out this sum explicitly, the subscripts on the individual forces remind us that every force is caused by another object. Further, that the forces we want to include are all the forces exerted by other objects on the object we are considering.

$$\vec{F}_A^{net} = \vec{F}_{B \to A} + \vec{F}_{C \to A} + \vec{F}_{D \to A} + ... = \sum_j \vec{F}_{j \to A}$$

5. *m* -- Dividing the net force by m (subscript A) reminds us that the resulting force on the object is shared over the parts of the object. A bigger object will have less of a response (acceleration) to the same force.

6. → --The little arrows on top of the acceleration and net force remind us that Newton's second law is a vector equation. This means that each perpendicular direction has its own Newton's law -- x, y, and z. Further, that it is the net force in the x direction that affects the motion in the x direction, the net force in the y direction that affects the motion in the y direction, etc.

That's a lot to pack into one little equation with what looks like 3 symbols (that turn out to be 6). But each of these ideas is an essential piece of making sense of this important principle and illustrates how much complex knowledge can be represented externally in what looks like a "simple" equation.

This kind of cognitive and conceptual packing into a mathematical representation is strongly imbedded in the culture of physics, even at the introductory level. From my recent interactions in the NEXUS project, this is less common in introductory chemistry and rare in introductory biology. As a result, our biology students may not be accustomed to this kind of knowledge coding and need some explicit help to get them beyond framing equations as purely calculational tools.[40,41]

## 6. Conclusion

In this paper I have attempted to demonstrate the value that can be added to education research, development, and reform by taking a theoretical perspective. The overarching issues of context, cultural, representations, and their interaction show more structure when viewed in this way. And our improved understanding of this structure helps us to not oversimplify situations in which we might be first tempted to overlook their complexity at our peril.

## Acknowledgements

This paper is dedicated to my good friend and mentor, Leonard Jossem (1919-2009), who first introduced me to the international physics education community. Much of this paper has arisen from dis-cussions with members of the University of Maryland PERG and BERG and with the members of the NEXUS development team. I particularly thank Ben Dreyfus, Ben Geller, Kristi Hall, and Mike Hull for specific contributions and comments to this paper. This work was made possible by the support of the Howard Hughes Medical Institute's NEXUS Project and the US National Science Foundation Awards 09-19816 and DUE 11-22818. Any opinions, findings, and conclusions or recommendations expressed in this publication are those of the author(s) and do not necessarily reflect the views of the National Science Foundation.

# References

[1] E. F. Redish, "A Theoretical Framework for Physics Education Research: Modeling student thinking," in *Proceedings of the International School of Physics, "Enrico Fermi" Course CLVI*, Varenna, Italy, August 2003, E. F. Redish and M. Vicentini (eds.) (IOS Press, Amsterdam, 2004).

[2] E. F. Redish and K. A. Smith, "Looking Beyond Content: Skill development for engineers," *Journal of Engineering Education* 97 (July 2008) 295-307.

[3] Bibliography for "Resources: A Theoretical Framework"
[http://www.physics.umd.edu/perg/tools/ResourcesReferences.pdf]

[4] H. Roediger and K. McDermott, *Journal of Experimental Psychology: Learning, Memory, & Cognition*, 21 (1995) 803-814.

[5] The two test words are "needle" and "sleep".

[6] Bartlett, *Remembering: A Study in Experimental and Social Psychology* (Cambridge U. Press, 1995/1932).

[7] John Kotre, *White Gloves* (Norton, 1998).

[8] R. L. Buckner & D. C. Carroll, "Self-projection and the brain," *Trends in Cognitive Science* 11:2 (2006), 49-57.

[9] D. J. Simons & C. F. Chabris, "Gorillas in our midst: sustained inattentional blindness for dynamic events", *Perception* 28 (1999) 1059-1074; M. S. Ambinder & D. J. Simons, "Attention Capture: The interplay of expectations, attention, and awareness", in L. Itti et al., (eds.) *The Neurobiology of Attention* (Elsevier, 2005), 69-75.

[10] G. A. Miller, "The magical number seven, plus or minus two: Some limits on our capacity for processing information," *Psychological Review* 63 (1956) 81-97.

[11] A. Baddeley, *Human Memory: Theory and Practice, Revised Edition* (Allyn & Bacon, 1998).

[12] A. H. Johnstone, "Chemistry Teaching – Science or Alchemy?" *Journal of Chemical Education*, 74:3 (1997) 262-268.

[13] J. Fuster, *Cortex and Mind: Unifying cognition* (Oxford U. Press, 2003).

[14] M. Bar, et al., "Top-down facilitation of visual recognition," *Proc. Natl. Acad. Of Sci.*, 103:2 (Jan. 10, 2006) 449-454.

[15] N. Mesgarani & E. F. Chang, "Selective cortical representation of attended speaker in multi-talker speech perception," *Nature* (2012) doi:10.1038/nature11020.

[16] J. S. Brown & P. Duguid, *The Social Life of Information* (Harvard Bus. Rev. Press, 2000).

[17] E. Hutchins, *Cognition in the Wild* (Bradford Books, 1996).

[18] E. Goffmann, *Frame Analysis: An essay on the organization of experience* (Northeastern U. Press, 1986).

[19] D. Tannen, *Framing in Discourse* (Oxford U. Press, 1993).

[20] G. L. MacClachlan and I. Reid, *Framing and Interpretation* (Melbourne U. Press, 1994).

[21] D. Hammer, A. Elby, R. E. Scherr, & E. F. Redish, "Resources, framing, and transfer," in *Transfer of Learning: Research and Perspectives*, J. Mestre (ed.) (Information Age Publishing, 2004).

[22] B. Frank, *The Dynamics of Variability in Introductory Physics Students' Thinking: Examples from Kinematics*, PhD dissertation, Department of Physics, University of Maryland (2009).

[23] E. F. Redish and D. Hammer, Reinventing College Physics for Biologists: Explicating an Epistemological Curriculum , *Am. J. Phys.*, **77**, 629-642 (2009)

[24] L. C. McDermott, P. S. Shaffer, and the Physics Education Group at the University of Washington, *Tutorials in Introductory Physics*, (Prentice Hall, , 1998).

[25] E. F. Redish, *Teaching Physics with the Physics Suite* (John Wiley & Sons, Inc., 2003).

[26] A. A. diSessa, "Toward an Epistemology of Physics," *Cognition and Instruction*, 10 (1993)105-225.

[27] *Characteristics of Scientists and Engineers in the United States: 2006*, NSF report 11-318 (National Center for Science and Engineering Statistics, 2011).

[28] E. F. Redish, Adding value through interdisciplinary conversations, *plenary talk*, 2nd Conference on Transforming Research in Undergraduate STEM Education (TRUSE), St. Paul, MN, 3. June 2012.

[29] *Bio 2010: Transforming Undergraduate Education for Future Research Biologists* (National Academies Press, 2003).

[30] AAMC-HHMI committee, *Scientific Foundations for Future Physicians.* (AAMC, 2009)

[31] For addition reports on articles concerning the reform of biology education, see the NEXUS Documents on Biology Education Reform.

[32] Project NEXUS (HHMI) [http://www.hhmi.org/grants/office/nexus/]

[33] Project NEXUS (UMCP) [http://umdberg.pbworks.com/w/page/44091483/Project NEXUS UMCP]

[34] K. L. Hall, J. E. Watkins, J. E. Coffey, T. J. Cooke, and E. F. Redish, Examining the Impact of Student Expectations on Undergraduate Biology Education Reform, AERA 2011 report.

[35] W. C. Galley, "Exothermic bond breaking: A persistent misconception," *J. Chem. Ed.*, 81:4 (2004) 523-525.

[36] E. F. Redish, J. M. Saul, and R. N. Steinberg, On the Effectiveness of Active-Engagement Microcomputer-Based Laboratories, *Am. J. Phys.* **65**, 45-54 (January 1997).

[37] R. M. Goertzen, R. Scherr, & A. Elby, Accounting for tutorial teaching assistants' buy-in to reform instruction, *Phys. Rev. – Special Topics in PER*, 5, 020109 (2009) [20 pages]; "Tutorial teaching assistants in the classroom: Similar teaching behaviors are supported by varied beliefs about teaching and learning," *Phys. Rev. – Special Topics in PER*, 6, 010105 (2010) [17 pages]; "Respecting tutorial instructors' beliefs and experiences: A case study of a physics teaching assistant," *Phys. Rev. – Special Topics in PER*, 6, 020125 (2010) [11 pages].

[38] H. W. Stevenson and J. W. Stigler, *The Learning Gap: Why Our Schools Are Failing and What We Can Learn from Japanese and Chinese Education* (Summit Books, 1992).

[39] M. Hull, PhD dissertation, Physics Department, University of Maryland, expected 2012.

[40] J. Tuminaro and E. F. Redish, "Elements of a Cognitive Model of Physics Problem Solving: Epistemic Games," *Phys. Rev. – Special Topics in PER*, **3**, 020101 (2007).

[41] E. F. Redish, Problem Solving and the Use of Math in Physics Courses, paper from talk given at the conference *World View on Physics Education in 2005: Focusing on Change, Delhi, August 21-26, 2005.*